\documentclass{ieeeaccess}
\usepackage{cite}
\usepackage{amsmath,amssymb,amsfonts}
\usepackage{algorithmic}
\usepackage{booktabs}
\usepackage{color}
\usepackage{graphicx}
\usepackage{multirow}
\usepackage{textcomp}
\def\BibTeX{{\rm B\kern-.05em{\sc i\kern-.025em b}\kern-.08em
    T\kern-.1667em\lower.7ex\hbox{E}\kern-.125emX}}
    
\usepackage{caption}
\DeclareCaptionFont{ieeeblue}{\color{accessblue}}
\DeclareCaptionLabelFormat{myformat}{\figcapheadfont{#1 #2}}
\begin{document}
% \history{Date of publication xxxx 00, 0000, date of current version xxxx 00, 0000.}
\history{}
% \doi{10.1109/ACCESS.2017.DOI}

\title{Modeling of Rakugo Speech and Its Limitations: Toward Speech Synthesis That Entertains Audiences}
\author{\uppercase{Shuhei Kato\authorrefmark{1,2}},
\IEEEmembership{Student Member, IEEE},
\uppercase{Yusuke Yasuda\authorrefmark{1,2}},
\IEEEmembership{Student Member, IEEE},
\uppercase{Xin Wang\authorrefmark{2}},
\IEEEmembership{Member, IEEE},
\uppercase{Erica Cooper\authorrefmark{2}},
\IEEEmembership{Member, IEEE},
\uppercase{Shinji Takaki\authorrefmark{3}},
\IEEEmembership{Member, IEEE}
\uppercase{and Junichi Yamagishi\authorrefmark{2,4}},
\IEEEmembership{Senior Member, IEEE}}
\address[1]{The Graduate University for Advanced Sciences (SOKENDAI), Hayama, Kanagawa, Japan}
\address[2]{National Institute of Informatics, Chiyoda, Tokyo, Japan}
\address[3]{Nagoya Institute of Technology, Nagoya, Aichi, Japan}
\address[4]{The University of Edinburgh, Edinburgh, UK}
\tfootnote{This work was partially supported by a JST CREST Grant (JPMJCR18A6, VoicePersonae project), Japan, and by MEXT KAKENHI Grants (16H06302, 17H04687, 18H04120, 18H04112, 18KT0051), Japan. The authors would like to thank Prof.\ Masaru Kitsuregawa of the National Institute of Informatics for kindly providing useful suggestions on rakugo speech synthesis.}

\markboth
{Kato \headeretal: Modeling of Rakugo Speech and Its Limitations: Toward Speech Synthesis That Entertains Audiences}
{Kato \headeretal: Modeling of Rakugo Speech and Its Limitations: Toward Speech Synthesis That Entertains Audiences}

\corresp{Corresponding author: Shuhei Kato (e-mail: skato@nii.ac.jp).}

\begin{abstract}
We have been investigating \textit{rakugo} speech synthesis as a challenging example of speech synthesis that entertains audiences. Rakugo is a traditional Japanese form of verbal entertainment similar to a combination of one-person stand-up comedy and comic storytelling and is popular even today. In rakugo, a performer plays multiple characters, and conversations or dialogues between the characters make the story progress. To investigate how close the quality of synthesized rakugo speech can approach that of professionals' speech, we modeled rakugo speech using Tacotron 2, a state-of-the-art speech synthesis system that can produce speech that sounds as natural as human speech albeit under limited conditions, and an enhanced version of it with self-attention to better consider long-term dependencies. We also used global style tokens and manually labeled context features to enrich speaking styles. Through a listening test, we measured not only naturalness but also distinguishability of characters, understandability of the content, and the degree of entertainment. Although we found that the speech synthesis models could not yet reach the professional level, the results of the listening test provided interesting insights: 1) we should not focus only on the naturalness of synthesized speech but also the distinguishability of characters and the understandability of the content to further entertain audiences; 2) the fundamental frequency ($f_{o}$) expressions of synthesized speech are poorer than those of human speech, and more entertaining speech should have richer $f_{o}$ expression. Although there is room for improvement, we believe this is an important stepping stone toward achieving entertaining speech synthesis at the professional level.
\end{abstract}


\titlepgskip=-15pt

\maketitle

\section{Introduction}
\label{sec:introduction}
\PARstart{C}{an} machines read texts aloud like humans? The answer is yes, albeit under limited conditions. The mean opinion scores (MOSs) of some speech synthesis (text-to-speech; TTS) systems are the same as those of natural speech~\cite{ShenJ2018,LiN2019}. These systems are trained with well-articulated read speech. Attempts to model speech with various speaking styles have also been actively investigated in deep-learning-based speech synthesis studies~\cite{WangY2017uncovering,WangY2018,SkerryRyanRJ2018,WoodT2018,HsuWN2019,VasquezS2019,BattenburgE2019,SunG2020}.

However, can machines verbally entertain like humans? The answer is probably no. Verbal entertainment, including \textit{rakugo}, a traditional Japanese form of verbal entertainment similar to a combination of one-person stand-up comedy and comic storytelling, entertains audiences through the medium of speech. In other words, speech in verbal entertainment does not just transfer information such as content and speaker emotions, personality, and intention to listeners, but also stirs listeners' emotions. Most of us would agree that verbal-entertainment performances by machines are quite unnatural or monotonic even if the content is appropriate. Some speech-synthesis-based rakugo performances, incorporating many manual interventions, have been submitted to online video platforms~\cite{MSS2009,MetsukiWaruiP2011,zky2012}. You might enjoy such performances depending on your opinion, but we believe these performances have far poorer quality than those by professional rakugo performers. We believe such a gap between machines and humans should be filled to evolve human-machine relationships. However, to the best of our knowledge, no speech synthesis studies have directly estimated how large such a gap is. In this study, we directly measured the degree of entertainment of synthesized speech and human speech through a listening test. If there is a gap, the results of the listening test will suggest how speech synthesis for entertaining audiences should be improved.

Before modeling rakugo speech, we first recorded and built a large rakugo speech database for our experiments because most commercial rakugo recordings are live recordings that contain noise and reverberation; thus, no rakugo speech databases suited to speech synthesis are available.

Using this database, we modeled rakugo speech with two speech synthesis systems, Tacotron 2~\cite{ShenJ2018} and an enhanced version of it with self-attention~\cite{ChengJ2016}. We believe Tacotron 2 is appropriate for our purpose because it is a state-of-the-art speech synthesis system that can produce (read-aloud) speech that sounds as natural as human speech, and can also effectively model audiobook speech, which is usually expressive speech like rakugo speech\footnote{The differences between rakugo speech and audiobook speech are introduced in \ref{subsec:difference_between_rakugo_speech_and_audiobook_speech}.}.~\cite{WangY2018,SkerryRyanRJ2018,HsuWN2019,ZenH2019}. In order to take prosodic characteristics of Japanese speech into consideration, we further enhance Tacotron 2 with self-attention (SA-Tacotron) for modeling rakugo speech because it is reported that the combination of Tacotron 1\footnote{The architectures of Tacotron 1~\cite{WangY2017tacotron} and Tacotron 2~\cite{ShenJ2018} are quite similar. We chose Tacotron 2 in this paper because the MOSs of synthesized speech from Tacotron 2 are the same as those of human speech while those of Tacotron are not.}~\cite{WangY2017tacotron} and self-attention outperforms Tacotron 1 in 
%pitch-accent languages such as 
Japanese~\cite{YasudaY2019investigation}. We also combined global style tokens (GSTs)~\cite{WangY2018} and/or manually labeled context features with Tacotron 2 and SA-Tacotron to enrich the speaking styles of synthetic rakugo speech.

If rakugo speech synthesis is possible, for example, we can enjoy a deceased performer's performance of a story that he or she never performed. Synthesized rakugo performance at a sufficient level may enable us to examine why \textit{humans} perform rakugo (and other forms of verbal entertainment) and what value \textit{only humans} can provide to audiences. It may consequently evolve rakugo performance itself.

We describe the continuation of research begun in our previous study~\cite{KatoS2019rakugo}. The differences from that study~\cite{KatoS2019rakugo} and this one are as follows: in this study, 1) the implementation of the backbone Tacotron 2 architecture in all the experimental systems (\ref{subsec:models_and_samples_used_in_experiments}) were revised to be consistent with the original Tacotron 2 paper~\cite{ShenJ2018}, except for the kind of the attention module (\ref{subsec:Tacotron2} and \ref{subsec:SA-Tacotron}) , 2) SA-Tacotron was used, 3) speech samples used in the following listening test were synthesized from models based on a systematic selection: Tacotron 2 or SA-Tacotron, with or without the combination of GSTs and/or manually labeled context features, and 4) a more detailed listening test was conducted involving directly asking listeners to answer how well they were entertained.

The rest of this paper is organized as follows: In Section II, we give an overview of rakugo. In Section III, we give the details of our rakugo speech database. In Section IV, we explain Tacotron 2 and introduce SA-Tacotron. In Section V, we present the test conditions, results, and discussion. We conclude the paper and mention future work in Section VI.

\section{Rakugo}

\subsection{Overview}

\Figure[t](topskip=0pt, botskip=0pt, midskip=0pt)[scale=0.23]{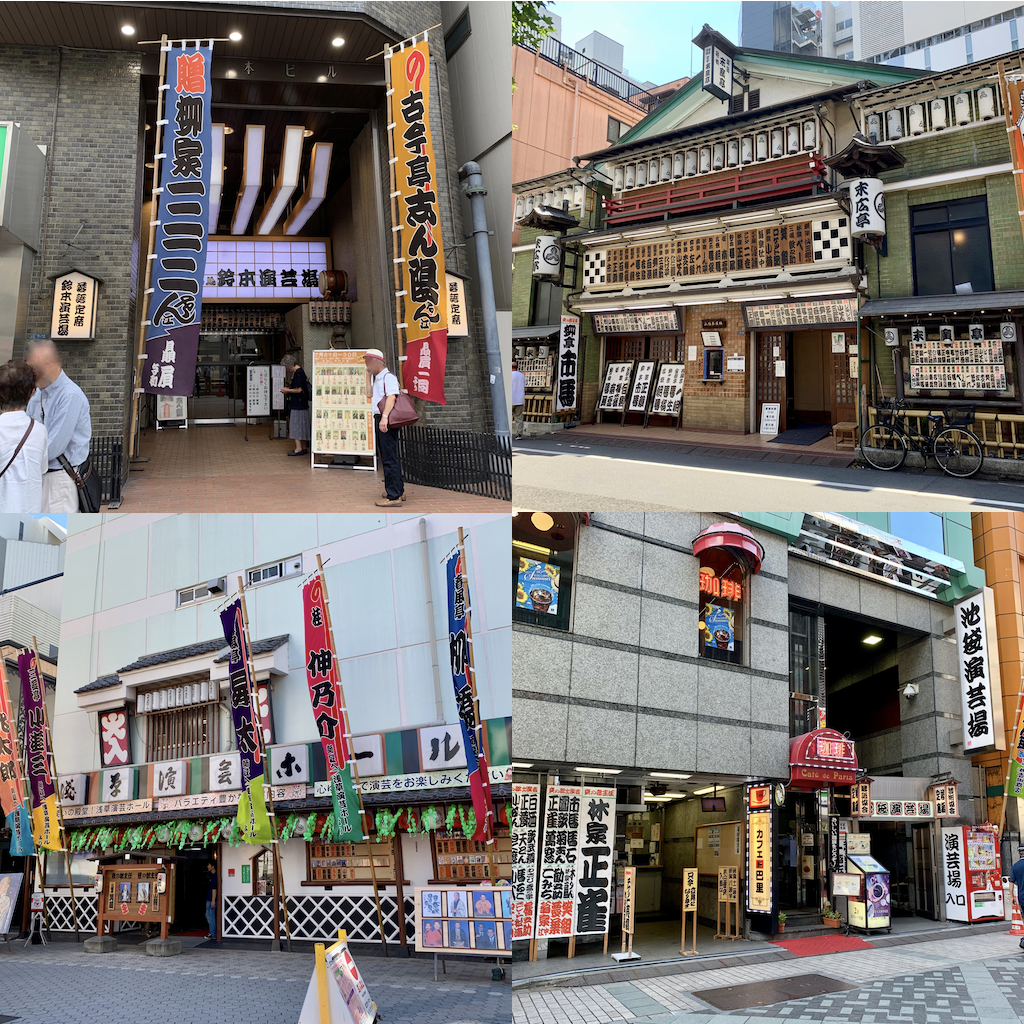}{Four major yoses (theaters mainly producing rakugo) in Tokyo. Upper left: Suzumoto Engeijo~\cite{SuzumotoEngeijo}, upper right: Suehirotei~\cite{Suehirotei}, lower left: Asakusa Engei Hall~\cite{AsakusaEngeiHall}, and lower right: Ikebukuro Engeijo~\cite{IkebukuroEngeijo}.\label{fig:yose}}

Rakugo has over 300 years of history, and it is a popular form of entertainment even today in Japan. A theater that mainly produces rakugo is called a \textit{yose}. In Tokyo, there are four major yoses, and rakugo is performed in each one every day of the year, even on January 1 (Figure~\ref{fig:yose}). There are dozens of other minor yoses. Rakugo is also performed at small and large halls, restaurants, coffeehouses, bookstores, shrines, temples, etc.\ almost every day. Thousands of CDs, DVDs, and streaming audio and videos of rakugo performances by current or former professional rakugo performers are available. Some TV and radio programs are broadcast every week in Japan~\cite{YoseChannel,AsakusaOchanomaYose,KamigataRakugoNoKai,ShinuchiKyoen,ShinosukeRadio,YoseApuri,RadioYose}. Amateur performances are also active. Some amateur rakugo performance societies at universities have produced professional performers.

Rakugo is generally divided into \textit{Edo} (Tokyo) rakugo, which we focus on in this paper, and \textit{Kamigata} rakugo, which was developed in Osaka and Kyoto. A professional rakugo performer is called a \textit{hanashika}. In Edo rakugo, a hanashika is ranked at one of three levels, i.e., \textit{zenza} (minor performer, stage assistant, and housekeeper at their master/mistress's house), \textit{futatsume} (second-rank performer), and \textit{shin-uchi} (first-rank performer). Only shin-uchis can take disciples. Usually, it takes about 3 to 5 years to be promoted from zenza to futatsume and about 10 years to be promoted from futatsume to shin-uchi. About 600 performers are active as professionals in Edo rakugo as of 2020.

\subsection{Performance}

During a performance, a rakugo performer sits down on a \textit{zabuton} (cushion) and performs improvisationally or from memory alone on a stage (Figure~\ref{fig:Shumputei_Shotaro}). He or she plays multiple characters, and their conversations and dialogues make the story progress. Most of the main part of a story consists of conversations and dialogues between the characters played by the performer. In Edo rakugo, performers use only a \textit{sensu} (folding fan) and a \textit{tenugui} (hand towel) as props.

\Figure[t](topskip=0pt, botskip=0pt, midskip=0pt)[scale=0.2]{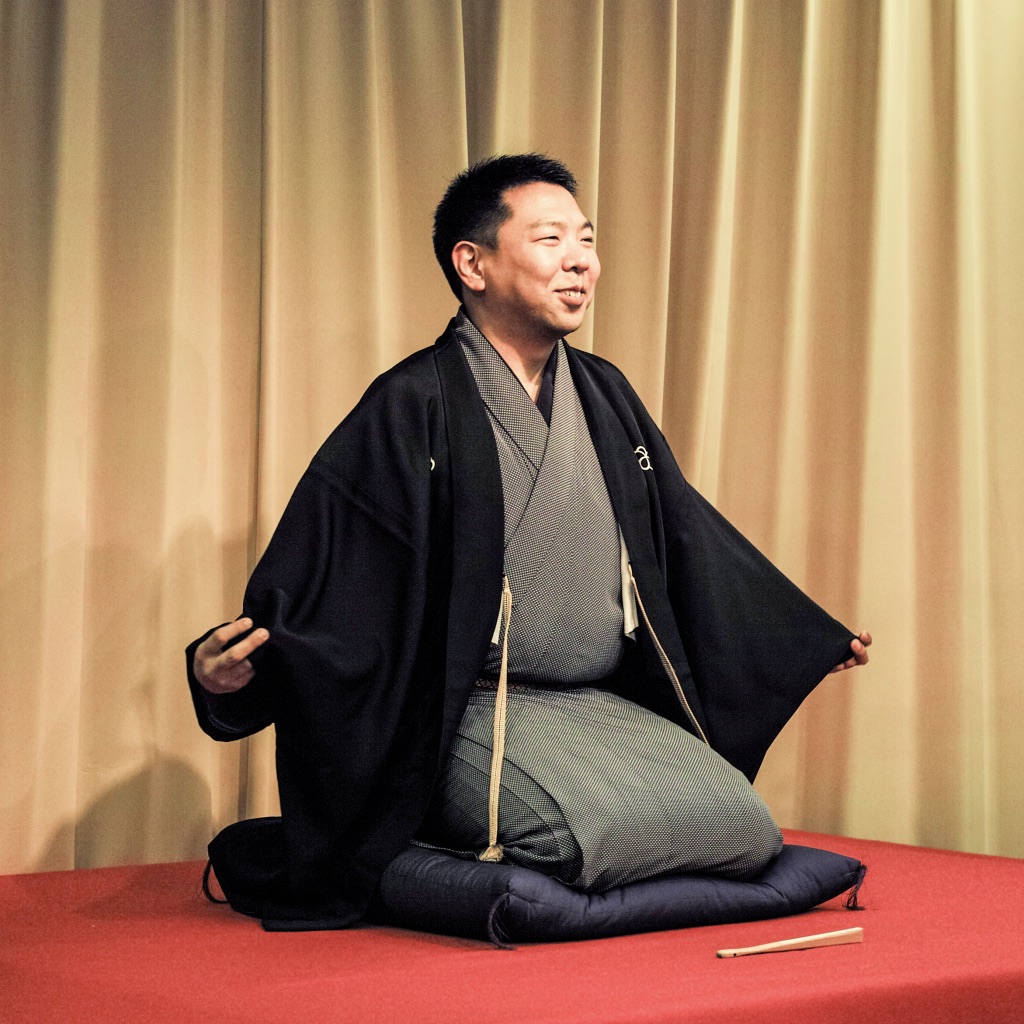}{Shumputei Shotaro~\cite{ShumputeiShotaro}, a professional rakugo performer, performing rakugo on a stage~\cite{akirakawamura2015}.\label{fig:Shumputei_Shotaro}}

Rakugo performers tell stories. A rakugo story is composed of five parts: \textit{maeoki} (greeting), \textit{makura} (introduction), the main part, \textit{ochi} (punch line), and \textit{musubi} (conclusion)~\cite{NomuraM1994}. Maeoki is optional, so it may not appear during a performance. As will be mentioned below, some exceptional stories do not have ochi. Such stories have musubi in place of ochi. Musubi is also used when performers terminate stories because of time limitations. Makura is often improvised, but during this, performers basically do not have conversations with the audience, unlike stand-up comedy\footnote{Of course, audiences are allowed to react nonverbally to the performance.}. Ochi is the most important part of rakugo (the word ``rakugo'' is derived from ``a story with ochi'').

Rakugo stories are divided into two groups depending on the time of establishment~\cite{YamamotoS2012}. A story established by about the 1910s is called a standard, and a story created after the 1920s is called a modern story. In this paper, we focus on standards. It should be noted that the Japanese language used in standards is slightly old-fashioned, and each character speaks a different dialect, sociolect, and idiolect of Japanese according to his or her gender, age, social rank, and individuality.

Rakugo stories can be divided into two genres depending on the contents~\cite{YamamotoS2012}. The more fundamental one is called \textit{otoshibanashi} or \textit{kokkeibanashi} (funny stories), which concentrates on making audiences laugh. Otoshibanashi has ochi at the end of the story. The other one is called \textit{ninjobanashi} (stories of humanity), which concentrates on making audiences impressed. Some ninjobanashi do not have ochi. In this paper, we focus on otoshibanashi.

The length of rakugo stories varies from story to story. Even if performers perform the same story, the length can vary because of time limitations or other situations. In a yose, one hanashika usually performs for about 15 minutes (only the last performer performs for about 30 minutes). In other stages or recordings, they may perform longer.

Rakugo stories are taught through oral instruction from a master or mistress to a disciple except when the story is new. Performers may edit stories to increase the quality or match their own characteristics. They sometimes insert jokes not only in the makura but also in the main parts of the stories according to the situations during their performances.

The following is an example of a very short rakugo paragraph (otoshibanashi).

\vspace{10pt}

\begin{quote}
{\sffamily Tome}: Whoa! Oh no! Oh no! Oh no! Oh no!\\
{\sffamily Friend}: Wait, Tome. What are you doing?\\
{\sffamily Tome}: Oh, I'm chasing after a thief.\\
{\sffamily Friend}: Seriously? Aren't you the fastest man in this town? He is unlucky.\\
{\sffamily Friend}: Which direction did he escape?\\
{\sffamily Tome}: He's catching up with me.
\end{quote}

\subsection{Difference between rakugo and audiobook speech/TTS}
\label{subsec:difference_between_rakugo_speech_and_audiobook_speech}

Some readers may wonder how rakugo speech and its TTS differ from the speech of audiobooks, which is an active research topic in the speech synthesis field, and its TTS. The main difference is that almost all parts of a rakugo story consist of conversations and dialogues between characters that are played by a performer from memory, and the conversations and dialogues make the story progress. As mentioned above, there are few narrative sentences in the conversational parts. In other words, rakugo performers should communicate the story to audiences without explicit explanations.

It should also be noted that rakugo speech is more casually pronounced than that of audiobooks because rakugo is performed from memory. In addition, expression of rakugo speech is far more diverse than that of audiobooks because the entire story is mostly comprised only of conversations and dialogues between characters. Also, as mentioned above, the Japanese language used in traditional rakugo stories is somewhat old-fashioned, and each character speaks a different dialect, sociolect, and idiolect according to his or her gender, age, social rank, and individuality.

Moreover, since rakugo is a form of entertainment, we argue that whether audiences are being entertained by rakugo speech or its TTS is essentially important. In addition, since the rakugo performance mainly consists of conversations and dialogues, building an appropriate rakugo TTS would also help us design improved conversational or acting speech synthesizers.

\section{Database}

\subsection{Overview}
\label{subsec:database_overview}

Although we can easily purchase commercial rakugo recordings from the market, most of them are not suitable for speech synthesis because of the room reverberation, environment noise, and sounds made by the audiences. Therefore, we recorded and built a rakugo speech database for our experiments.

The recordings were conducted from July to September 2017. The performer was Yanagiya Sanza~\cite{YanagiyaSanza}, a shin-uchi rakugo performer with over 20 years of professional experience. Only he was in the recording booth, and he did not face or receive any reaction from an audience (Figure~\ref{fig:Sanza_recording}). He performed 25 Edo rakugo standards, lasting from 6 to 47 minutes (total 13.2 hours including pauses between sentences). In order to record performances with as natural a flow as possible, we did not re-record any of the stories when mispronunciation or restatements happened, except in cases where the performer asked us to do so.

\Figure[t](topskip=0pt, botskip=0pt, midskip=0pt)[scale=0.26]{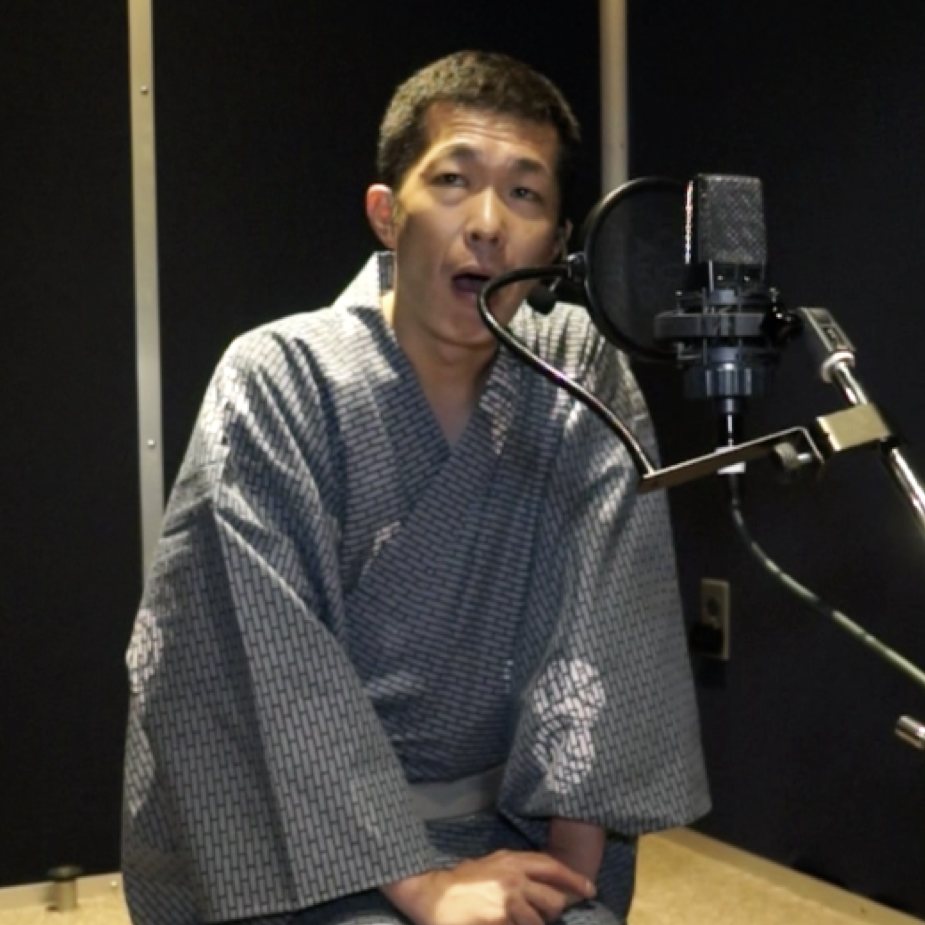}{Yanagiya Sanza performing rakugo alone in recording booth.\label{fig:Sanza_recording}}

The first author carefully transcribed the pronunciation of the recorded speech. We did not define any special symbols for mispronunciation, fillers, or laughs. We used a comma only as a pause in a sentence, a period at the end of a sentence, and a question mark at the end of a sentence that ends with rising intonation. The ratio of question sentences to the other sentences was about 3:7. We separated sentences according to the following criteria:

\begin{itemize}
    \item A place we can separate sentences grammatically followed by a pause.
    \item A place where turn-taking occurs.
    \item A place right after a rising intonation.
\end{itemize}

All the symbols used in the transcription are listed in Table~\ref{tab:symbols}. Although Japanese is a pitch-accent language~\cite{MoriGM1929,ChibaT1935,DanielsFJ1958,WenckG1966}, we did not use any accent symbols because the results of automatic morphological analysis and accent annotation on the slightly old-fashioned Japanese dialects in the rakugo speech are not accurate enough for speech synthesis tasks. Of course, manual labeling of accents is impractical because it is time consuming.

\begin{table}[t]
    \caption{Symbols used in transcription of rakugo database.}
    \label{tab:symbols}
    \centering
    \begin{tabular}{lll}
        \toprule
        \multirow{6}{*}{Phonemes} & Vowels & a, e, i, o, u\\
        \cmidrule{2-3}
        & \multirow{3}{*}{Consonants} & b, by, ch, d, dy, f, fy, g, gw, gy, h, hy, j, k,\\
        & & kw, ky, m, my, n, N, ng, ny, p, py, r, ry, s,\\
        & & sh, t, ts, ty, v, w, y, z\\
        \cmidrule{2-3}
        & Other & cl (geminate consonant)\\
        \cmidrule{1-3}
        \multirow{2}{*}{Pauses} & & pau (comma), sil (start of a sentence and\\
        & & period), qsil (question mark)\\
        \bottomrule
    \end{tabular}
\end{table}

\subsection{Context labels}

We also appended context labels to each sentence (Table~\ref{tab:context_lables}). All the labels, excluding {\bfseries part} of the story, were defined by the first author because no well-known categories of them exist in rakugo.

We believe the {\bfseries role} of the character is important because almost all speech in rakugo, especially in the main part, is composed of conversations or dialogues. The {\bfseries individuality} of the character is a special category for fool characters, who are usually called \textit{Yotaro} and often appear in rakugo stories. We believe the {\bfseries condition} of characters is also important because characters speak in various styles. All the styles were defined by the first author from carefully listening to speech and reading context. The {\bfseries relationship} of the talking companions was defined because in conversations or dialogues (between two characters) in rakugo, one must be considered the superior and the other the inferior. The {\bfseries n\_companion} (number of talking companions) was defined because characters may talk to themselves or speak to one person or multiple people. The {\bfseries distance} to the talking companions was introduced for conversations between characters. Some characters speak louder and more clearly as if they are speaking to someone far away. In the context of a particular {\bfseries part} of the story, we considered maeoki (greetings) and musubi (conclusion) as makura (introduction) and ochi (punch line), respectively.

\begin{table*}[t]
    \caption{Context labels (\textit{\sffamily hanashika} refers to speech not by any character).}
    \label{tab:context_lables}
    \centering
    \begin{tabular}{lll}
        \toprule
        {\bfseries Group} & {\bfseries Name} & {\bfseries Details}\\
        \midrule
        \multirow{2}{*}{{\bfseries ATTR}ibute of character} & {\bfseries role} of character & \parbox[c]{9cm}{{\bfseries gender}: \textit{hanashika}, male, female; {\bfseries age}: \textit{hanashika}, child, young, middle-aged, old; {\bfseries social rank}: \textit{hanashika}, {\itshape samurai}, artisan, merchant, other townsperson, countryperson, with other dialect, modern, other}\\
        \addlinespace[1.5mm]
        & {\bfseries individuality} of character & \textit{hanashika}, fool\\
        \midrule
        {\bfseries COND}ition of character & {\bfseries condition} of character & \parbox[c]{9cm}{neutral, admiring, admonishing, affected, angry, begging, buttering up, cheerful, complaining, confident, confused, convinced, crying, depressed, drinking, drunk, eating, encouraging, excited, feeling sick, sleepy, feeling sorry, finding it easier than expected, freezing, frustrated, ghostly, happy, hesitating, interested, justifying, {\itshape kakegoe} (shouting/calling), loud voice, laughing, leaning on someone, lecturing, looking down, panicked, pet-directed speech, playing dumb, putting up with, rebellious, refusing, sad, scared, seducing, shocked, shouting, sketchy, small voice, soothing, straining, surprised, suspicious, swaggering, teasing, telling off, tired, trying to remember, underestimating, unpleasant}\\
        \midrule
        \multirow{4.5}{*}{{\bfseries SIT}uation of character} & \parbox[c]{4cm}{{\bfseries relationship} to talking companions} & \textit{hanashika}, narrative, soliloquy, superior, inferior\\
        \addlinespace[1.5mm]
        & \parbox[c]{4cm}{{\bfseries n\_companion}: number of talking companions} & \textit{hanashika}, narrative, soliloquy, one, two or more\\
        \addlinespace[1.5mm]
        & {\bfseries distance} to talking companions & \textit{hanashika}, narrative, near, middle, far\\
        \midrule
        {\bfseries STR}ucture of story & {\bfseries part} of story & makura, main part, ochi\\
        \bottomrule
    \end{tabular}
\end{table*}

\section{Tacotron 2/SA-Tacotron-based rakugo speech synthesis}

\subsection{Why is end-to-end TTS needed?}

We first tried traditional pipeline (frame-by-frame) neural speech synthesis models to synthesize rakugo speech, but the quality of synthesized speech was very poor{\footnote{Speech samples of pipeline models are available at {\ttfamily https://nii-\\yamagishilab.github.io/samples-rakugo/pipeline/}}. We therefore selected end-to-end (sequence-to-sequence) neural models.  There are two strong reasons underpinning the use of end-to-end models. The first reason is that automatic phoneme segmentation and automatic fundamental frequency ($f_{o}$) extraction are difficult for rakugo speech, especially for highly expressive speech. Pipeline models normally require phoneme boundary information as inputs~\cite{ZenH2013}.
%In most cases, we automatically estimate them and correct some of them. 
But in the case of rakugo speech, the result of automatic estimation was extremely poor. In addition, $f_{o}$ extraction also failed for non-speech sounds given by the performer, such as coughs, yawns, snores, sighs, and knocks, which sometimes play an important role in rakugo performance.}

The second reason is that rakugo speech uses slightly old-fashioned Japanese dialects as mentioned in \ref{subsec:database_overview}, and automatic morphological analysis and pitch-accent annotation defined for modern standard Japanese do not work properly although they are normally used in Japanese pipeline models as inputs~\cite{LuongHT2018}.

End-to-end TTS models, on the other hand, can be built without marking the phoneme boundaries, conducting the morphological analysis, or annotating the pitch-accent labels. The $f_{o}$ extraction can also be avoided if the model predicts a mel-spectrogram or waveform. Therefore, end-to-end TTS models are more suitable for rakugo speech.

\subsection{Tacotron 2}
\label{subsec:Tacotron2}

Tacotron 2~\cite{ShenJ2018} is a state-of-the-art end-to-end speech synthesis system that produces read-aloud speech that sounds as natural as human speech. Some Tacotron-based systems can effectively model expressive speech including audiobook speech~\cite{WangY2018,SkerryRyanRJ2018,HsuWN2019,ZenH2019}. We therefore argue that modeling rakugo speech using Tacotron 2 is reasonable.

The architecture of a Tacotron-2-based rakugo TTS model is shown in Figure~\ref{fig:Tacotron2}. This model is slightly modified from the original Tacotron 2 to learn alignments more robustly and quickly.

\Figure[t](topskip=10pt, botskip=0pt, midskip=10pt)[scale=0.5]{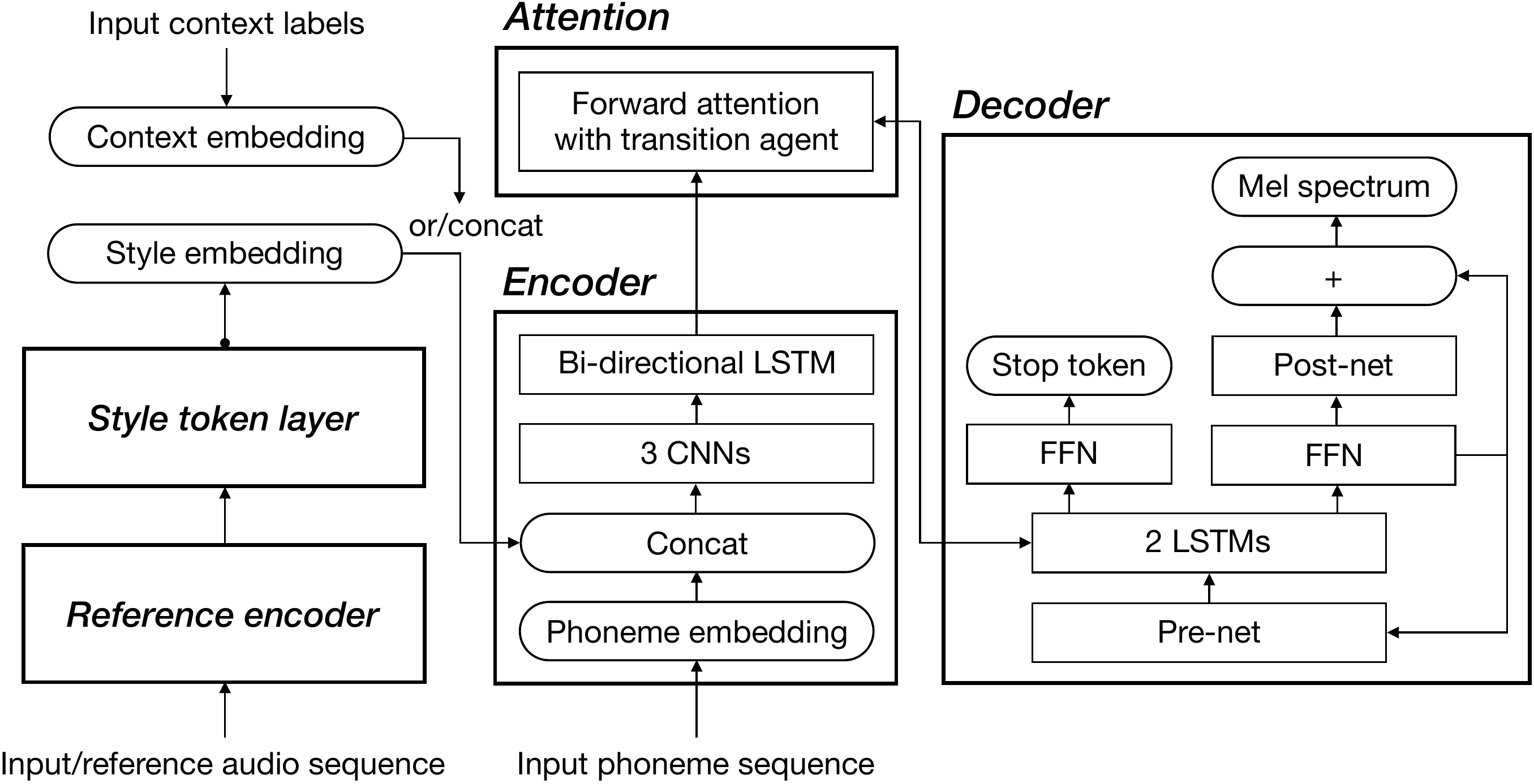}{Overall network structure of Tacotron-2-based rakugo TTS. Network structure of reference encoder and style token layer is shown in Figure~\ref{fig:GST}.\label{fig:Tacotron2}}

The main network is composed of an encoder, decoder, and attention network, which maps the time steps of the encoder to those of the decoder. The encoder converts an input phoneme sequence into a hidden feature representation, which will be taken to predict an output mel spectrogram. Each input phoneme is embedded into a 512-dimensional vector. The vector can then be concatenated to a style embedding, which is described in \ref{subsec:GST}, a context embedding, which is derived from the input context labels listed in Table~\ref{tab:context_lables}, or a concatenated vector of them. The (concatenated) vector is passed into 3 convolutional neural networks (CNNs) each containing 512 filters with a $5 \times 1$ shape (same padding~\cite{DumoulinV2016}), followed by batch normalization~\cite{IoffeS2015} and rectified linear unit (ReLU) activations. The output of the final CNN is then passed into a single bi-directional~\cite{SchusterM1997} long short-term memory (LSTM)~\cite{HochreiterS1997} that has 512 units (256 units per direction) to generate the final encoded features.

The output sequence of the encoder is used by the attention network, which compresses the full encoded sequence as a fixed-length context vector for each decoder time step. We used forward attention with a transition agent~\cite{ZhangJX2018} instead of the location sensitive attention~\cite{ChorowskiJ2015} used in the original Tacotron 2 to learn the alignment between the encoder and decoder time steps more robustly and quickly. The forward-attention algorithm has a left-to-right initial alignment, which is useful because all the alignments of encoder-decoder TTS should be left-to-right since the output speech has to be produced from the beginning to the end of the input text. For further details, please refer to a previous study~\cite{ZhangJX2018}.

In the decoder, the predicted mel spectrum in the previous time step is first passed into a pre-net, a feed-forward network (FFN) that has 2 fully connected layers of 256 ReLU units. The pre-net output and the context vector from the attention network at the previous time step are concatenated and passed into 2 unidirectional LSTMs each containing 1,024 units. Using the output sequence of the LSTMs and the encoder output, the context vector at the current time step is calculated. Then, the concatenation of the output of the LSTMs and the context vector is passed into an FFN that has a fully connected layer of 80 linear units to generate a mel spectrum. To predict a residual for improving the reconstruction, the predicted spectrum is passed into a post-net, which consists of 5 CNNs each containing 512 filters with a $5 \times 1$ shape (same padding) followed by batch normalization. Tanh activations are then applied except for the final layer. The summation of the former predicted mel spectrum and the output of the post-net is the final target mel spectrum.

In parallel with the spectrogram prediction, the concatenation of the output of the decoder LSTMs and the attention vector is projected to a scalar and activated by a sigmoid function to predict the probability of the completion of the output sequence. This probability is called the ``stop token.''

Training is conducted through minimizing the summation of the mean squared errors (MSEs) of the spectrograms both before and after the post-net, the binary cross-entropy loss of the stop token, and the L2 regularization loss. During training, dropout~\cite{SrivastavaN2014} is applied to each layer of the FFN in the pre-net and the CNNs in the post-net with probability 0.5 for regularization. Zoneout~\cite{KruegerD2017} is also applied to each LSTM with probability 0.1 for further regularization.

\subsection{SA-Tacotron}
\label{subsec:SA-Tacotron}

For further improvement, we enhanced the Tacotron 2 above with self-attention~\cite{ChengJ2016} (SA-Tacotron). Self-attention can effectively capture long-term dependencies, and it was reported that the enhanced version of Tacotron~\cite{WangY2017tacotron} with self-attention exceeds Tacotron for Japanese~\cite{YasudaY2019investigation}. We therefore enhanced the Tacotron 2 above with self-attention. The network structure of SA-Tacotron is shown in Figure~\ref{fig:SA-Tacotron}.

\Figure[t](topskip=10pt, botskip=0pt, midskip=10pt)[scale=0.5]{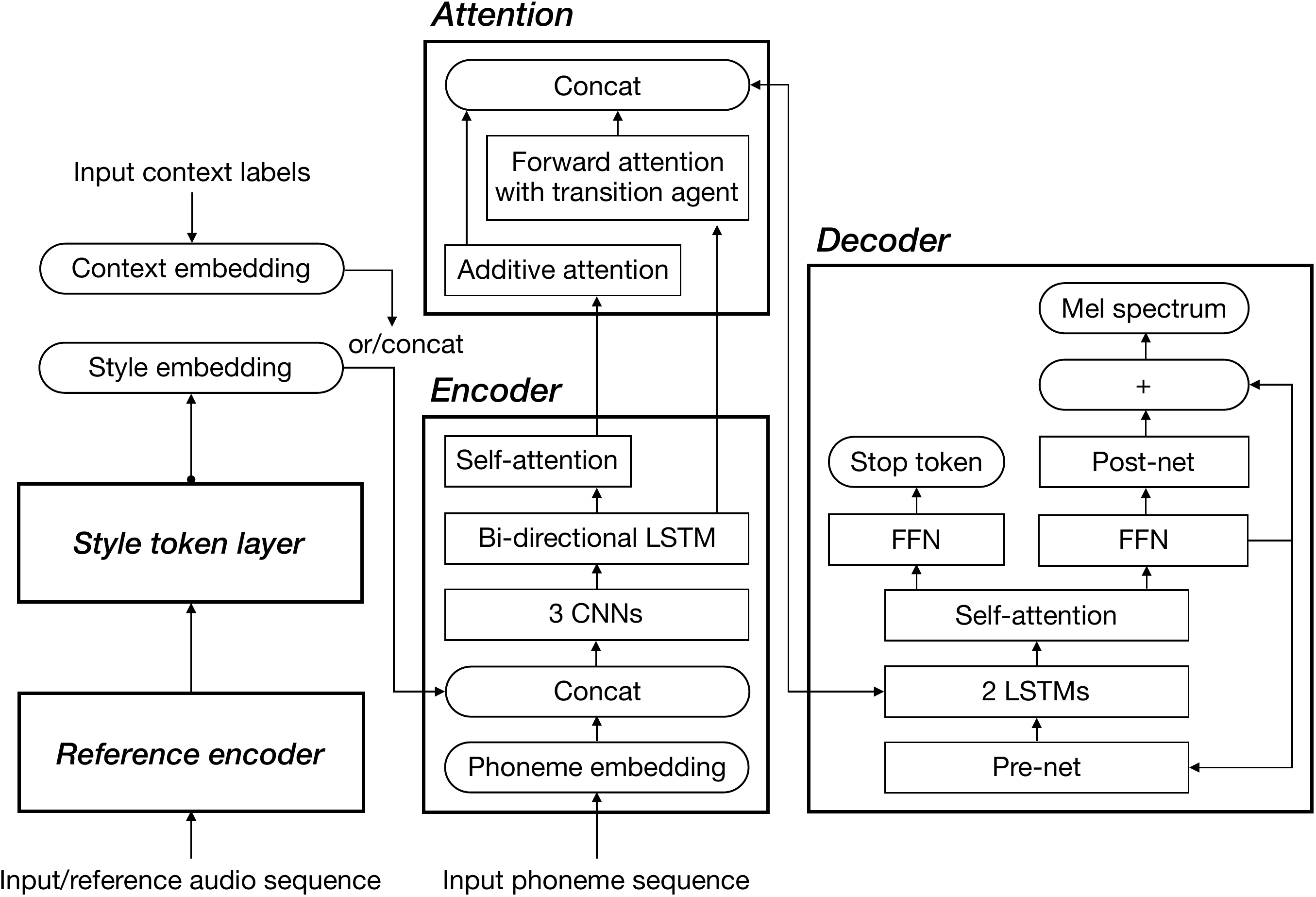}{Overall network structure of SA-Tacotron-based rakugo TTS. Network structure of reference encoder and style token layer is shown in Figure~\ref{fig:GST}.\label{fig:SA-Tacotron}}

In the encoder, a self-attention block is inserted after the bi-directional LSTM. A self-attention block consists of a self-attention network, followed by a fully connected layer with tanh activation and residual connections. This block is expected to capture the long-term dependencies inside the input phoneme sequence. The self-attention network is implemented with a 2-head multi-head attention~\cite{VaswaniA2017} mechanism. The self-attention block outputs a sequence of 32-dimensional vectors. During training, dropout is applied to the multi-head attention mechanism with probability of 0.05 for regularization. The encoder generates two output sequences, one is from the bi-directional LSTM and the other from the self-attention block.

The output sequences of the encoder are input into two attention networks. The output sequence from the bi-directional LSTM is used by a forward attention network with a transition agent, the same architecture as in Tacotron 2 described in \ref{subsec:Tacotron2}. On the other hand, the output sequence from the self-attention block is used by an additive attention~\cite{BahdanauD2015} network. The context vectors from the two attention networks are concatenated and used in the decoder.

The structure of the decoder is the same as that of the decoder of Tacotron 2 described in \ref{subsec:Tacotron2}, except that a self-attention block is inserted after the LSTMs. This block is expected to capture the long-term dependencies inside the output sequence. It used two attention heads, and its output has 1,568 dimensions ($1024 + 512 + 32$).

\subsection{Global style tokens with Tacotron 2 and SA-Tacotron}
\label{subsec:GST}

We also use GSTs~\cite{WangY2018} to enrich the speaking style of synthesized speech and make characters distinguishable from each other. The GST framework was proposed as a prosody transfer approach. In this framework, we assume that TTS systems have access to a reference audio file from which we can extract the prosody and speaking style and transfer them to the synthesized speech from the TTS system. The role of the GST is to extract the prosody and speaking style that cannot be explained by the text input.

\Figure[t](topskip=0pt, botskip=0pt, midskip=10pt)[scale=0.5]{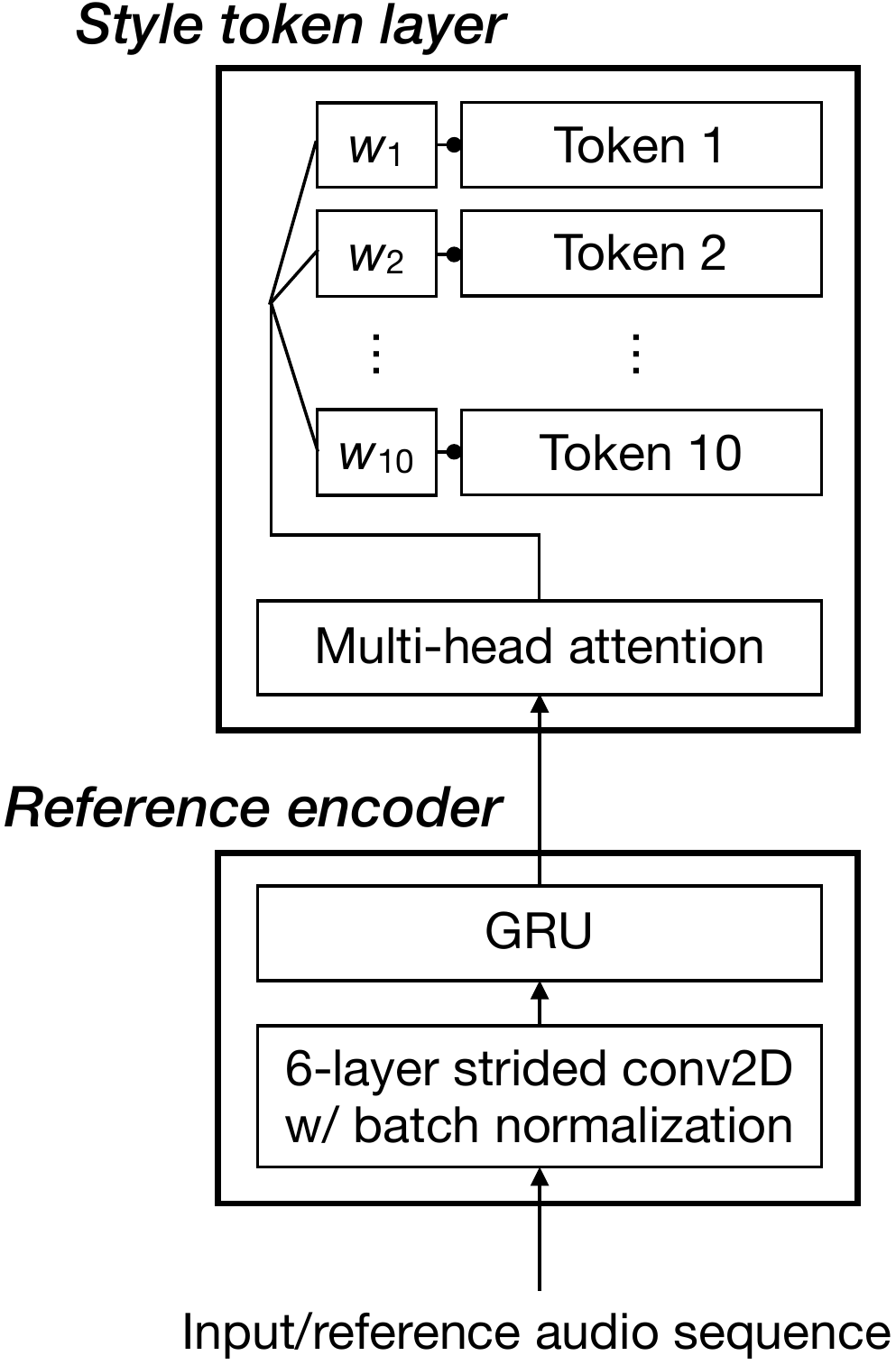}{Network structure of reference encoder and style token layer.\label{fig:GST}}

The architecture we used is basically the same as the original GST~\cite{WangY2018}, except for some parameters (Figure~\ref{fig:GST}). An input or reference audio sequence, i.e., 80-dimensional mel spectrogram, is passed into a reference encoder. The reference encoder is composed of 6-layer 2D convolution layers with batch normalization and a 128-unit gated recurrent unit (GRU)~\cite{ChoK2014}. Each convolution layer is made up of $3 \times 3$ filters with $2 \times 2$ stride (same padding) and ReLU activation. Batch normalization is applied to each layer. The number of filters in the layers are 128, 128, 256, 256, 512, and 512. The output of the final layer is passed into the GRU. The final state of the GRU is then passed into a network called a style token layer.

The style token layer is composed of 10 randomly initialized 512-dimensional embeddings called style tokens and a multi-head attention. The output of the reference encoder and tanh-activated tokens are then mapped by the multi-head attention network. Any number of attention heads can be used if the dimension of style tokens can be divided by this number. If the number of heads is $h$, the dimension of tokens is $512/h$. We used 8 heads in this study based on the results of preliminary experiments. The attention network calculates the weights over tokens, and the weighted summation of tokens is treated as a style embedding, which will be concatenated to the phoneme embeddings output from the encoder of the Tacotron-2-based or SA-Tacotron-based model. The style embedding vector is constant within a sentence.

\section{Listening test}

\subsection{Purpose of listening test}

We modeled rakugo speech with two different types of models, Tacotron-2-based (\ref{subsec:Tacotron2}) and SA-Tacotron-based (\ref{subsec:SA-Tacotron}). We conducted a listening test to compare their performances and assess how they were accepted by the public.

Since rakugo is a form of entertainment and the conversations or dialogues between the characters make the story progress, we are interested in not only the naturalness of synthesized speech but also how accurately listeners distinguish characters, how well listeners understand the content of the speech, and how well the speech entertains listeners.

\subsection{Models and samples used in listening tests}
\label{subsec:models_and_samples_used_in_experiments}

We used 16 fully-annotated stories out of the total of 25 stories in the database for our listening test\footnote{Annotation of the database is a work in progress.}. The 16 stories are about 4.3 hours long  in total, except for pauses between sentences, and contain 7,341 sentences. We used 6,437 sentences for training, 715 for validation, and 189 for testing. The training and validation sets did not include very short ($< 0.5$\,s) or very long ($\ge 20$\,s) utterances to reduce alignment errors during training.

We trained several models for the listening test.

\begin{itemize}
    \item {\bfseries\sffamily Tacotron} is a Tacotron-2-based model, and its input is the phoneme sequence. No style embeddings or context features were used as an input.
    \item {\bfseries\sffamily Tacotron-GST-8} is a Tacotron-2-based model including GSTs with an 8-head attention. It should be noted that the reference audio of the test set was natural speech\footnote{We used natural speech as the reference audio of the test set to compare *-GST-8 with other models (*-ATTR and *-context) that use manually labeled correct context features, which were labeled through listening to recorded speech of the test set, in a fair condition.}.
    \item {\bfseries\sffamily Tacotron-ATTR} is a Tacotron-2-based model with manually labeled context features belonging to ATTR (role and individuality) only. The dimension of the context embedding is 4.
    \item {\bfseries\sffamily Tacotron-context} is a Tacotron-2-based model with all of the manually labeled context features. The dimension of the context embedding is 68.
    \item {\bfseries\sffamily Tacotron-GST-8-ATTR} and {\bfseries\sffamily Tacotron-GST-8-con-} {\bfseries\sffamily text} are Tacotron-2-based models with a combination of GSTs with an 8-head attention and manually labeled context features belonging to ATTR or all contexts, respectively.
    \item {\bfseries\sffamily SA-Tacotron}, {\bfseries\sffamily SA-Tacotron-GST-8}, {\bfseries\sffamily SA-Tacotron-ATTR}, {\bfseries\sffamily SA-Tacotron-context}, {\bfseries\sffamily SA-Tacotron-GST-8-ATTR}, and {\bfseries\sffamily SA-Tacotron-GST-8-context} are the same models as {\sffamily Tacotron}, {\sffamily Tacotron-GST-8}, {\sffamily Tacotron-ATTR}, {\sffamily Tacotron-context}, {\sffamily Tacotron-GST-8-ATTR}, and {\sffamily Tacotron-GST-8-context}, respectively, but they are based on SA-Tacotron instead of Tacotron 2.
\end{itemize}

We trained each model for about 2,000 epochs. The mini-batch size was 128. The initial learning rate was 0.001, and the learning rate was decayed exponentially. The optimization method was Adam, and the number of mel filters for input spectrograms was 80. The spectrograms were generated from $48$\,kHz/$16$\,bit waveforms with $50$-ms-long frame, $12.5$-ms frame shift, Hann window, and 4,096-long fast Fourier transform. Values of the spectrograms were normalized into 0 mean and 1 standard deviation at each dimension over all the data.

Predicted mel spectrograms were converted into waveforms by using a WaveNet~\cite{OordA2016} vocoder~\cite{WangX2018,TamamoriA2017} trained with natural mel spectrograms and waveforms of all the training, validation, and test sets\footnote{We used natural mel spectrograms and waveforms of not only the training and validation sets but also the test set for WaveNet training. This design is to make comparison with AbS speech fairer. It should be noted that we also compared the above WaveNet to one without the test set and made sure that there was no perceptual difference.}. The sampling rate of the output waveform was $16$\,kHz\footnote{The sampling rate of the output waveform was 16\,kHz although that of the predicted mel spectrograms was 48\,kHz. This is because of computational complexity of the WaveNet model. Speech samples are available at {\ttfamily https://nii-yamagishilab.github.io/samples-rakugo/\\201910\_IEEE\_access/}}.

\subsection{Test conditions}

\Figure[t](topskip=0pt, botskip=0pt, midskip=0pt)[scale=0.27]{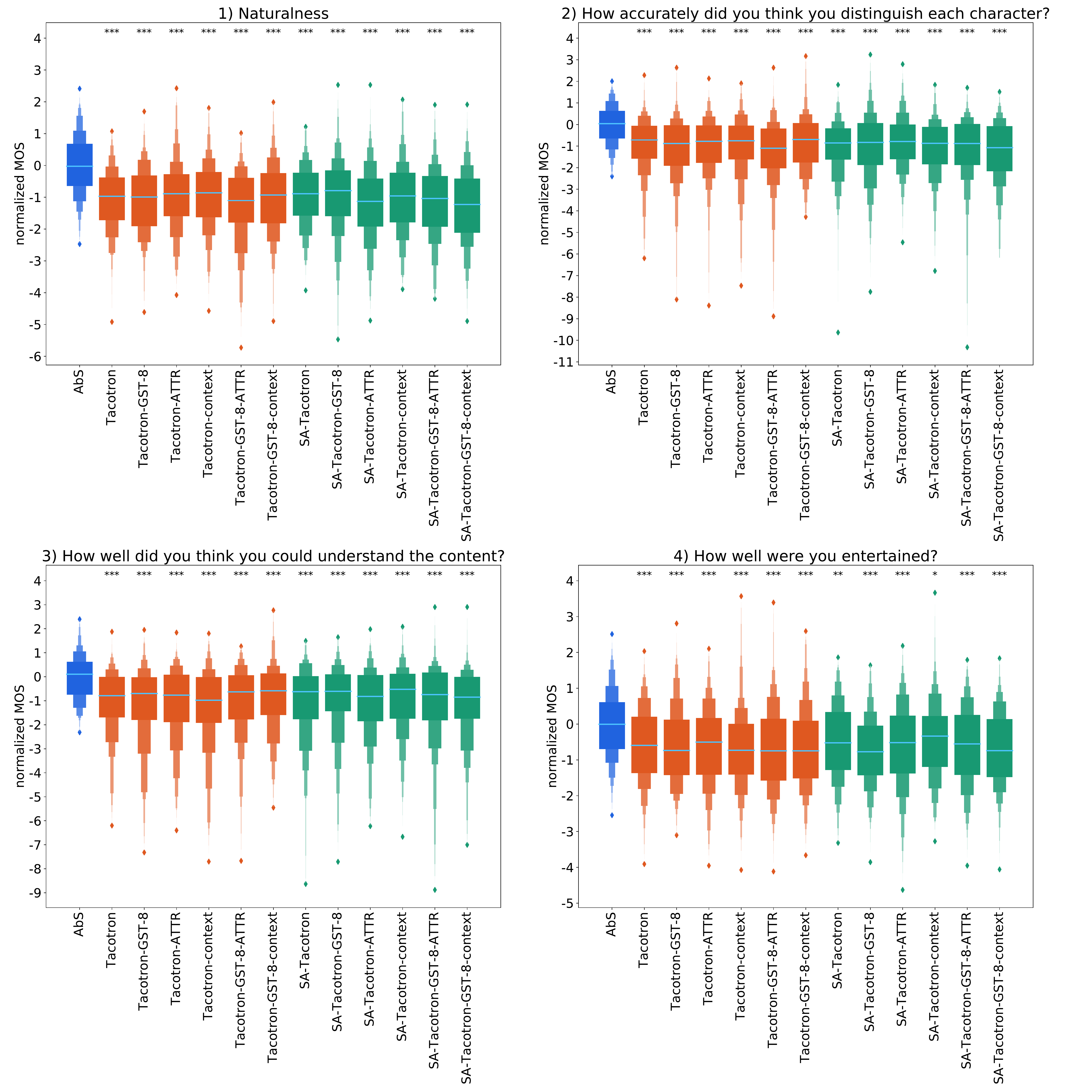}{Boxen plots for each question of listening test. Light blue lines represent medians. *: $p < 0.01$, **: $p < 0.005$, ***: $p < 0.001$. Only significant differences via Brunner-Munzel test with Bonferroni correction between AbS speech and each (SA-)Tacotron-based model are shown. There were no significant differences among (SA-)Tacotron-based models.\label{fig:result_listening_test}}

We selected a set of sentences comprising short stories as materials for the listening test. A total of 189 sentences comprising 13 short stories were prepared, and sentence-level audio files were concatenated as one audio file per story. The duration of the stories lasts from 11 seconds to 1 minute and 58 seconds (total 11 minutes and 14 seconds) in the case of real recordings. Because the speech samples were predicted sentence by sentence, and pauses between sentences were not predicted, the pauses between sentences used in the test were the same as those of real audio recordings. In other words, pauses between sentences were taken from real speech, and other prosody including intra-sentence pauses were predicted using models. It is obvious that pauses between sentences should also be predicted using models, but that is out of the scope of this paper. Listeners evaluated speech NOT sentence by sentence but in a whole story. Analysis-by-synthesis (AbS; copy synthesis) speech was also used for the test. The concatenated audio files were normalized to $-26$\,dBov by sv56~\cite{ITU2005}.

We used MOS as the metric for the test. In each evaluation round, listeners listened to the speech of all 13 stories each synthesized using one of the models listed in \ref{subsec:models_and_samples_used_in_experiments} or AbS speech. For each listener, the story-system combinations and their permutation were randomly selected. The audio of one story was presented on each screen, and listeners answered four MOS-based questions:

\vspace{10pt}

\begin{description}
    \item[1)] How natural did the speech sound? (naturalness) 
    \item[2)] How accurately did you think you could distinguish each character?
    \item[3)] How well did you think you could understand the content?
    \item[4)] How well were you entertained?
\end{description}

\vspace{10pt}

We used a five-point MOS scale. A listener was allowed to answer only one evaluation round because listeners would remember the content of the stories. A total of 183 listeners participated in 183 evaluation rounds.

\subsection{Results}
\label{subsec:result_listening_test}

For fair comparison, the scores were 1) normalized to 0 mean and 1 standard deviation for each listener then 2) further normalized per story so that the mean score of the AbS of human performance would be 0 and the standard deviation of it would be 1. Operation 1 absorbs variations of scores among listeners, and operation 2 diminishes the effects of the content of the story. 

The results are shown in Figure~\ref{fig:result_listening_test}. For statistical analysis, we conducted a Brunner-Munzel test~\cite{Brunner2000} with Bonferroni correction among the (normalized) scores for all the model combinations. For Q1--Q3, AbS speech was superior to all the (SA-)Tacotron-based models. Regarding Q4, AbS speech was also superior to all these models, but the differences between some models and AbS speech were smaller. We could not find any significant differences among (SA-)Tacotron-based models.

We compared the results systematically for Tacotron and SA-Tacotron, with and without GST and/or context labels, but we found no significant trends.

\subsection{Discussion}

As mentioned in \ref{sec:introduction}, Tacotron 2 can produce speech that sounds as natural as human speech in the case of well-articulated read speech. However, in the case of rakugo speech, all the TTS models including Tacotron 2 could not achieve the same scores regarding naturalness as that for AbS speech. Regarding distinguishability of characters and understandability of content, there were also significant differences between the scores for each model and AbS speech. In other words, speech synthesis currently cannot reach the professional level of rakugo performance. 

The most important point of our listening test was whether the listeners were entertained by synthetic rakugo speech. However, none of the TTS models obtained scores equivalent to AbS speech. Interestingly, there were lower significant differences between the scores for some of the TTS models and AbS speech.

\Figure[t](topskip=0pt, botskip=0pt, midskip=0pt)[scale=0.6]{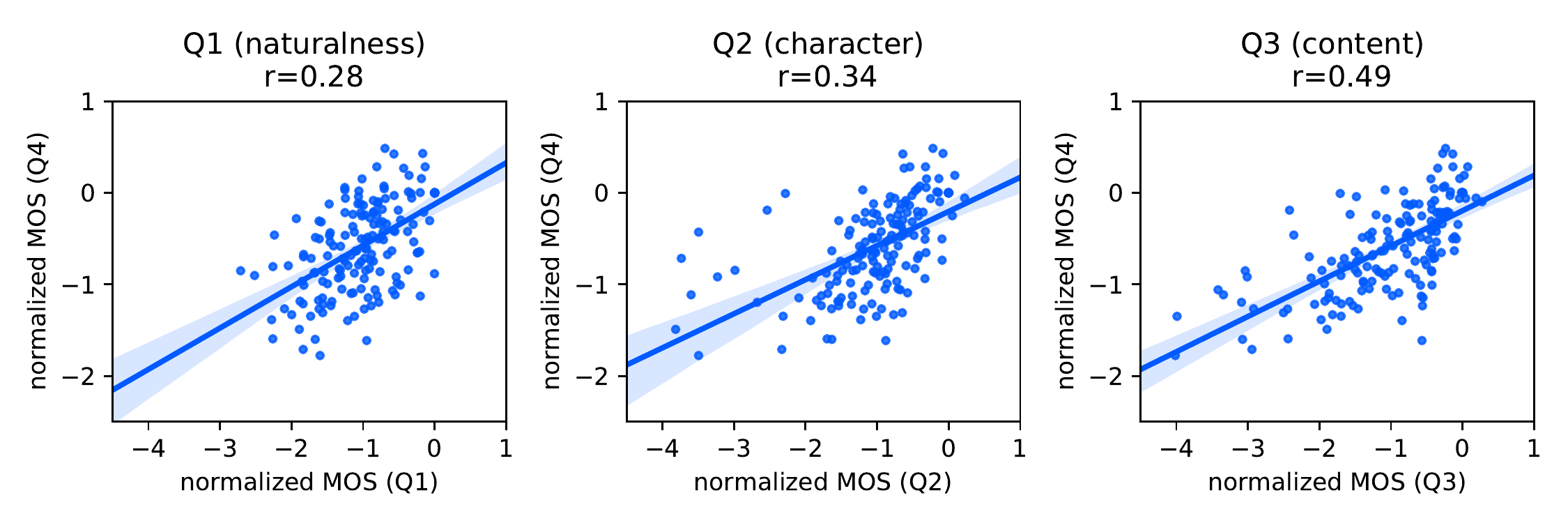}{Scatter plots and correlation coefficients of normalized MOSs between each of Q1--Q3 and Q4. Blue lines and light blue areas represent simple linear regression lines and 95\% confidence intervals, respectively.\label{fig:correlation_of_scores}}

\Figure[t](topskip=0pt, botskip=0pt, midskip=0pt)[scale=0.58]{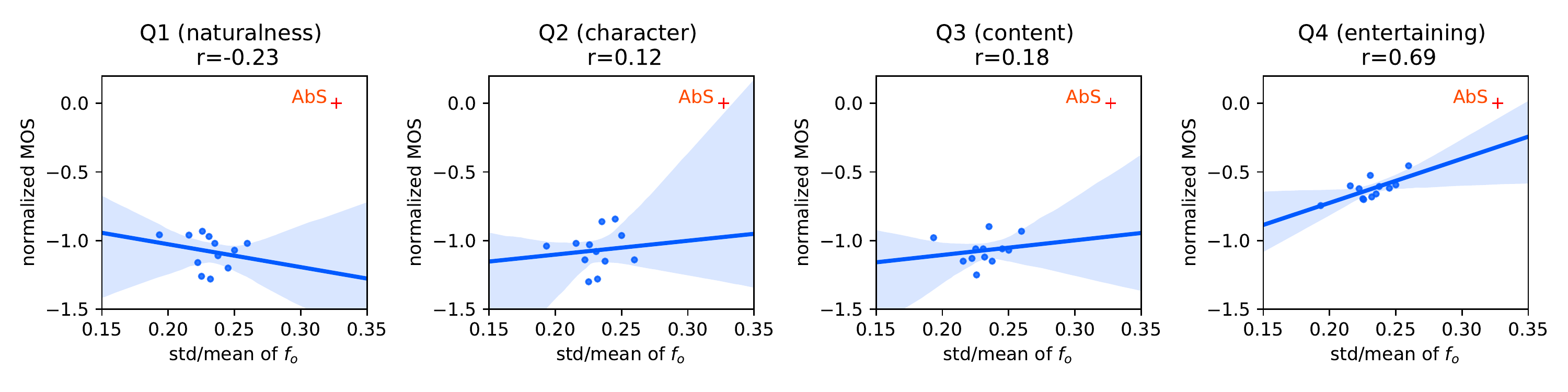}{Scatter plots and correlation coefficients of ratio of standard deviation to mean of $f_{o}$ and normalized MOSs. Blue lines and light blue areas represent simple linear regression lines and 95\% confidence intervals, respectively. AbS (red plus marks) were not considered when calculating correlation coefficients, regression lines, and confidence intervals.\label{fig:correlation_of_stdmean_fo_and_scores}}

For further analysis, we calculated the correlation among the scores for Q4, the question for evaluating the degree of entertainment, and those for the other questions. The results are shown in Figure~\ref{fig:correlation_of_scores}. The correlation coefficients between the scores for Q4 and those for Q2 (distingishability of characters) and between the scores for Q4 and those for Q3 (understandability of content) were higher than the coefficient between the scores for Q4 and those for Q1 (naturalness). This suggests that we should not only focus on the naturalness of synthesized speech, but we should also aim to improve the distinguishability of characters and the understandability of the content in order to further entertain listeners. We believe that this is an important insight for the speech synthesis community since speech synthesis research has thus far mainly focused on the naturalness over other aspects.

What should be improved to further entertain listeners in particular? To investigate this, we analyzed the relationship between $f_{o}$ and speech rates of the natural speech and of each model, and the scores for each model. We extracted $f_{o}$ for all the voiced frames ($5$-ms frame shift) in each sentence. The $f_{o}$s were extracted by WORLD~\cite{MoriseM2016} and then corrected manually. The extracted $f_{o}$s were concatenated over all the test sentences per model, and the means and standard deviations were calculated.

The relationship between the ratio of the standard deviation of $f_{o}$ to its mean and the scores of each question that we evaluated are shown in Figure~\ref{fig:correlation_of_stdmean_fo_and_scores}. We can clearly see that only Q4 (entertaining) has moderate correlation between the $f_{o}$'s variations and its scores. This suggests that more entertaining speech should have richer $f_{o}$ expression.

Speech rate is defined as the ratio of the number of mora to the duration of the speech. The means and standard deviations were calculated over all the test sentences per model. The relationship between the ratio of the standard deviation of speech rate to its mean and the scores are shown in Figure~\ref{fig:correlation_of_stdmean_speech_rates_and_scores}. Unlike the case of $f_{o}$, Q4 (entertaining) does not seem to correlate with the speech rate variations. Q1 (naturalness) and Q2 (distingishability of characters) have similar trends. Q3 (understandability of content) seems to have a negative correlation with the speech rate variations. However, its slope is almost flat.

\Figure[t](topskip=0pt, botskip=0pt, midskip=0pt)[scale=0.58]{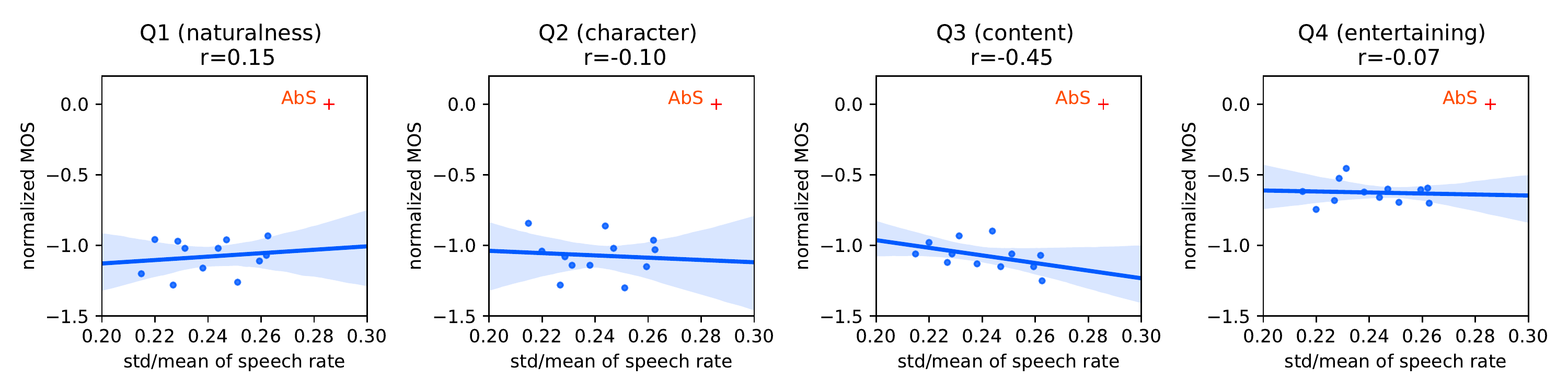}{Scatter plots and correlation coefficients of the ratio of standard deviation to mean of speech rates and normalized MOSs. Blue lines and light blue areas represent simple linear regression lines and 95\% confidence intervals, respectively. AbS (red plus marks) was not considered when calculating correlation coefficients, regression lines, and confidence intervals.\label{fig:correlation_of_stdmean_speech_rates_and_scores}}

\section{Conclusion}

Toward speech synthesis that entertains audiences, we modeled rakugo speech using Tacotron 2 and an enhanced version of it with self-attention (SA-Tacotron) to better consider long-term dependencies, and compared their outputs. We first built a rakugo database because no suitable databases or commercial recordings existed. We then trained TTS models with Tacotron 2 and SA-Tacotron. We also used global style tokens (GSTs) and manually labeled context features to enrich speaking styles. Through a listening test, however, we found that  state-of-the-art TTS models could not reach the professional level, and there were statistically significant differences in terms of naturalness, distinguishability of characters, understandability of content, and even the degree of entertainment; nevertheless, the results of the listening test provided some interesting insights: 1) we should not focus only on naturalness of synthesized speech but also the distinguishability of characters and the understandability of the content to further entertain listeners; 2) the $f_{o}$ expressivity of synthesized speech is poorer than that of human speech, and more entertaining speech should have richer $f_{o}$ expression.

We believe that this is an important step toward realization of genuinely entertaining TTS. However, many new components are needed for a fully automated rakugo synthesizer: 
\begin{itemize}
    \item Pauses between sentences need to be modeled; 
    \item GSTs need to be predicted from input texts and/or manually labeled context features;
    \item Fillers and filled pauses need to be automatically inserted into transcriptions;
    \item Only short stories were used in this study; much longer stories need to be synthesized and evaluated. 
\end{itemize}
In addition, there are open questions beyond simple TTS tasks such as:   
\begin{itemize}
    \item How can a rakugo synthesizer use reactions and feedback from an audience? 
    \item Can we generate rakugo stories automatically? 
    \item Can we create an audiovisual rakugo performer?
\end{itemize}
We hope that this paper becomes a seed for such new research directions. 

\bibliographystyle{IEEEtran}
\bibliography{mybib}

% Generated by IEEEtran.bst, version: 1.13 (2008/09/30)
\begin{thebibliography}{10}
\providecommand{\url}[1]{#1}
\csname url@samestyle\endcsname
\providecommand{\newblock}{\relax}
\providecommand{\bibinfo}[2]{#2}
\providecommand{\BIBentrySTDinterwordspacing}{\spaceskip=0pt\relax}
\providecommand{\BIBentryALTinterwordstretchfactor}{4}
\providecommand{\BIBentryALTinterwordspacing}{\spaceskip=\fontdimen2\font plus
\BIBentryALTinterwordstretchfactor\fontdimen3\font minus
  \fontdimen4\font\relax}
\providecommand{\BIBforeignlanguage}[2]{{%
\expandafter\ifx\csname l@#1\endcsname\relax
\typeout{** WARNING: IEEEtran.bst: No hyphenation pattern has been}%
\typeout{** loaded for the language `#1'. Using the pattern for}%
\typeout{** the default language instead.}%
\else
\language=\csname l@#1\endcsname
\fi
#2}}
\providecommand{\BIBdecl}{\relax}
\BIBdecl

\bibitem{ShenJ2018}
J.~Shen, R.~Pang, R.~J. Weiss, M.~Schuster, N.~Jaitly, Z.~Yang, Z.~Chen,
  Y.~Zhang, Y.~Wang, R.~Skerry-Ryan, R.~A. Saurous, Y.~Agiomyrgiannakis, and
  Y.~Wu, ``Natural {TTS} synthesis by conditioning {WaveNet} on mel spectrogram
  predictions,'' in \emph{Proc. International Conference on Acoustics, Speech,
  and Signal Processing (ICASSP)}, Calgary, AB, Canada, Apr 15--20 2018, pp.
  4779--4783.

\bibitem{LiN2019}
N.~Li, S.~Liu, Y.~Liu, S.~Zhao, and M.~Liu, ``Neural speech synthesis with
  transformer network,'' in \emph{Proc. AAAI Conference on Artificial
  Intelligence (AAAI-19)}, Honolulu, HI, USA, Jan 27 -- Feb 1 2019.

\bibitem{WangY2017uncovering}
Y.~Wang, R.~Skerry-Ryan, Y.~Xiao, D.~Stanton, J.~Shor, E.~Battenberg, R.~Clark,
  and R.~A. Saurous, ``Uncovering latent style factors for expressive speech
  synthesis,'' in \emph{Proc. Conference on Neural Information Processing
  Systems (NIPS) Machine Learning for Audio Signal Processing Workshop}, Long
  Beach, CA, USA, Dec 4--9 2017, pp. 4006--4010.

\bibitem{WangY2018}
Y.~Wang, D.~Stanton, Y.~Zhang, R.~Skerry-Ryan, E.~Battenberg, J.~Shor, Y.~Xiao,
  F.~Ren, Y.~Jia, and R.~A. Saurous, ``Style tokens: Unsupervised style
  modeling, control and transfer in end-to-end speech synthesis,'' in
  \emph{Proc. International Conference on Machine Learning (ICML)}, Stockholm,
  Stockholm, Sweden, Jul 10--15 2018, pp. 5180--5189.

\bibitem{SkerryRyanRJ2018}
R.~Skerry-Ryan, E.~Battenberg, Y.~Xiao, Y.~Wang, D.~Stanton, J.~Shor, R.~J.
  Weiss, R.~Clark, and R.~A. Saurous, ``Towards end-to-end prosody transfer for
  expressive speech synthesis with {Tacotron},'' arXiv:1803.09047 [cs.CL], Mar
  24 2018.

\bibitem{WoodT2018}
\BIBentryALTinterwordspacing
T.~Wood, ``Varying speaking styles with neural text-to-speech,'' Alexa Blogs,
  Nov 19 2018. [Online]. Available:
  \url{https://developer.amazon.com/zh/blogs/alexa/post/7ab9665a-0536-4be2-aaad-18281ec59af8/varying-speaking-styles-with-neural-text-to-speech}
\BIBentrySTDinterwordspacing

\bibitem{HsuWN2019}
W.-N. Hsu, Y.~Zhang, R.~J. Weiss, H.~Zen, Y.~Wu, Y.~Wang, Y.~Cao, Y.~Jia,
  Z.~Chen, J.~Shen, P.~Nguyen, and R.~Pang, ``Hierarchical generative modeling
  for controllable speech synthesis,'' in \emph{Proc. International Conference
  on Learning Representations (ICLR)}, New Orleans, LA, USA, May 6--9 2019.

\bibitem{VasquezS2019}
S.~Vasquez and M.~Lewis, ``{MelNet}: A generative model for audio in the
  frequency domain,'' arXiv:1906.01083 [eess.AS], Jun 4 2019.

\bibitem{BattenburgE2019}
E.~Battenburg, S.~Mariooryad, D.~Stanton, R.~Skerry-Ryan, M.~Shannon, D.~Kao,
  and T.~Bagby, ``Effective use of variational embedding capacity in expressive
  end-to-end speech synthesis,'' arXiv:1906.03402 [cs.CL], Jun 8 2019.

\bibitem{SunG2020}
G.~Sun, Y.~Zhang, R.~J. Weiss, Y.~Cao, H.~Zen, A.~Rosenberg, B.~Ramabhadran,
  and Y.~Wu, ``Generating diverse and natural text-to-speech samples using a
  quantized fine-grained {VAE} and auto-regressive prosody prior,'' in
  \emph{Proc. International Conference on Acoustics, Speech, and Signal
  Processing (ICASSP)}, May 4--8 2020.

\bibitem{MSS2009}
\BIBentryALTinterwordspacing
MSS, ``{Hanashika Miku No Tokusen Rakugo Manju Kowai Desu (in Japanese)},'' Jan
  20 2009. [Online]. Available: \url{https://www.nicovideo.jp/watch/sm5899050}
\BIBentrySTDinterwordspacing

\bibitem{MetsukiWaruiP2011}
\BIBentryALTinterwordspacing
{Metsuki-warui-P}, ``{[VOCALOID Rakugo] Kamban No Pin [Metsuki Warui Miku] (in
  Japanese)},'' Mar 25 2011. [Online]. Available:
  \url{https://www.nicovideo.jp/watch/sm13959846}
\BIBentrySTDinterwordspacing

\bibitem{zky2012}
\BIBentryALTinterwordspacing
zky, ``{[Hatsune Miku] VOCALOID Rakugo ``Nozarashi'' (in Japanese)},'' Feb 24
  2012. [Online]. Available: \url{http://www.nicovideo.jp/watch/sm17066984}
\BIBentrySTDinterwordspacing

\bibitem{ChengJ2016}
J.~Cheng, L.~Dong, and M.~Lapata, ``Long short-term memory-networks for machine
  reading,'' in \emph{Proc. Conference on Empirical Methods in Natural Language
  Processing (EMNLP)}, Austin, TX, USA, Nov 1--5 2016, pp. 551--561.

\bibitem{ZenH2019}
H.~Zen, V.~Dang, R.~Clark, Y.~Zhang, R.~J. Weiss, Y.~Jia, Z.~Chen, and Y.~Wu,
  ``{LibriTTS}: A corpus derived from {LibriSpeech} for text-to-speech,'' in
  \emph{Proc. INTERSPEECH}, Graz, Styria, Austria, Sep 15--19 2019, pp.
  1526--1530.

\bibitem{WangY2017tacotron}
Y.~Wang, R.~J. Skerry-Ryan, D.~Stanton, Y.~Wu, R.~J. Weiss, N.~Jaitly, Z.~Yang,
  Y.~Xiao, Z.~Chen, S.~Bengio, Q.~Le, Y.~Agiomyrgiannakis, R.~Clark, and R.~A.
  Saurous, ``Tacotron: Towards end-to-end speech synthesis,'' in \emph{Proc.
  INTERSPEECH}, Stockholm, Stockholm, Sweden, Aug 20--24 2017, pp. 4006--4010.

\bibitem{YasudaY2019investigation}
Y.~Yasuda, X.~Wang, and J.~Yamagishi, ``Investigation of enhanced {Tacotron}
  text-to-speech synthesis systems with self-attention for pitch accent
  language,'' in \emph{Proc. International Conference on Acoustics, Speech, and
  Signal Processing (ICASSP)}, Brighton, England, UK, Sep 12--17 2019, pp.
  6905--6909.

\bibitem{KatoS2019rakugo}
S.~Kato, Y.~Yasuda, X.~Wang, E.~Cooper, S.~Takaki, and J.~Yamagishi, ``Rakugo
  speech synthesis using segment-to-segment neural transduction and style
  tokens --- toward speech synthesis for entertaining audiences,'' in
  \emph{Proc. The 10th ISCA Speech Synthesis Workshop (SSW10)}, Vienna,
  Austria, Sep 20--22 2019, pp. 111--116.

\bibitem{SuzumotoEngeijo}
\BIBentryALTinterwordspacing
{Suzumoto Engeijo}, 7-12, Ueno 2-chome, Taito, Tokyo, Japan, 1857--present.
  [Online]. Available: \url{http://www.rakugo.or.jp}
\BIBentrySTDinterwordspacing

\bibitem{Suehirotei}
\BIBentryALTinterwordspacing
{Suehirotei}, 6-12, Shinjuku 3-chome, Shinjuku, Tokyo, Japan, 1897--present.
  [Online]. Available: \url{http://www.suehirotei.com}
\BIBentrySTDinterwordspacing

\bibitem{AsakusaEngeiHall}
\BIBentryALTinterwordspacing
{Asakusa Engei Hall}, 43-12, Asakusa 1-chome, Taito, Tokyo, Japan,
  1964--present. [Online]. Available: \url{http://www.asakusaengei.com}
\BIBentrySTDinterwordspacing

\bibitem{IkebukuroEngeijo}
\BIBentryALTinterwordspacing
{Ikebukuro Engeijo}, 23-1, Nishi-ikebukuro 1-chome, Toshima, Tokyo, Japan,
  1951--present. [Online]. Available: \url{http://www.ike-en.com}
\BIBentrySTDinterwordspacing

\bibitem{YoseChannel}
\BIBentryALTinterwordspacing
{Atoss Broadcasting Limited}, ``{Yose Channel},'' Oct 1 2012--present.
  [Online]. Available: \url{http://yosechannel.com}
\BIBentrySTDinterwordspacing

\bibitem{AsakusaOchanomaYose}
\BIBentryALTinterwordspacing
{Chiba Television Broadcasting Corporation}, ``{Asakusa Ochanoma Yose},'' Apr 5
  2004--present. [Online]. Available:
  \url{https://www.chiba-tv.com/program/detail/1012}
\BIBentrySTDinterwordspacing

\bibitem{KamigataRakugoNoKai}
\BIBentryALTinterwordspacing
{NHK}, ``{Kamigata Raukgo No Kai},'' Apr 24 2011--present. [Online]. Available:
  \url{https://www4.nhk.or.jp/P2851/}
\BIBentrySTDinterwordspacing

\bibitem{ShinuchiKyoen}
\BIBentryALTinterwordspacing
------, ``{Shin-uchi Kyoen},'' Nov 26 1978--present. [Online]. Available:
  \url{https://www4.nhk.or.jp/P632/}
\BIBentrySTDinterwordspacing

\bibitem{ShinosukeRadio}
\BIBentryALTinterwordspacing
{Nippon Cultural Broadcating Incorporated}, ``{Shinosuke Radio Rakugo De
  Date},'' Apr 7 2007--present. [Online]. Available:
  \url{http://www.joqr.co.jp/blog/rakugo/}
\BIBentrySTDinterwordspacing

\bibitem{YoseApuri}
\BIBentryALTinterwordspacing
{Nikkei Radio Broadcasting Corporation}, ``{Yose Apuri --- Warai Suginami Yose
  Kara},'' Oct 22 2017--present. [Online]. Available:
  \url{http://www.radionikkei.jp/yose/}
\BIBentrySTDinterwordspacing

\bibitem{RadioYose}
\BIBentryALTinterwordspacing
{TBS Radio Incorporated}, ``{Radio Yose},'' Oct 1974--present. [Online].
  Available: \url{https://www.tbsradio.jp/yose/}
\BIBentrySTDinterwordspacing

\bibitem{ShumputeiShotaro}
\BIBentryALTinterwordspacing
{Shumputei Shotaro}, Aug 23 1981--present. [Online]. Available:
  \url{http://shoutarou.com}
\BIBentrySTDinterwordspacing

\bibitem{akirakawamura2015}
{This photo is transformed from ``DP3M2471'' by akira kawamura licensed under
  CC BY 2.0}.

\bibitem{NomuraM1994}
M.~Nomura, \emph{Rakugo No Gengogaku (in Japanese)}.\hskip 1em plus 0.5em minus
  0.4em\relax Heibonsha, May 18 1994.

\bibitem{YamamotoS2012}
S.~Yamamoto, \emph{Rakugo No Rirekisho -- Kataritugarete 400-Nen (in
  Japanese)}.\hskip 1em plus 0.5em minus 0.4em\relax Shogakukan, Oct 6 2012.

\bibitem{YanagiyaSanza}
\BIBentryALTinterwordspacing
{Yanagiya Sanza}, Jul 4 1974--present. [Online]. Available:
  \url{http://www.yanagiya-sanza.com (in Japanese)}
\BIBentrySTDinterwordspacing

\bibitem{MoriGM1929}
M.~G. Mori, \emph{The Pronunciation of Japanese}.\hskip 1em plus 0.5em minus
  0.4em\relax The Herald-Sha, 1929.

\bibitem{ChibaT1935}
T.~Chiba, \emph{A study of accent: Research into its nature and scope in the
  light of experimental phonetics}.\hskip 1em plus 0.5em minus 0.4em\relax
  Fuzanbo Publishing Company, 1935.

\bibitem{DanielsFJ1958}
F.~J. Daniels, \emph{The Sound System of Standard Japanese: A Tentative Account
  from the Teaching Point of View with a Discussion of the `Accent' Theory
  (Accompanied by a Full Translation of a Japanese Account of the Theory) and
  of Word-Stress (Including a Consideration of Some Japanese Acoustical
  Work)}.\hskip 1em plus 0.5em minus 0.4em\relax Kenkyusha, 1958.

\bibitem{WenckG1966}
G.~Wenck, \emph{The Phonemics of Japanese: Questions and Attempts}.\hskip 1em
  plus 0.5em minus 0.4em\relax Harrassowitz, 1966.

\bibitem{ZenH2013}
H.~Zen, A.~Senior, and M.~Schuster, ``Statistical parametric speech synthesis
  using deep neural networks,'' in \emph{Proc. International Conference on
  Acoustics, Speech, and Signal Processing (ICASSP)}, Vancouver, BC, Canada,
  May 26--31 2013, pp. 7962–--7966.

\bibitem{LuongHT2018}
H.-T. Luong, X.~Wang, J.~Yamagishi, and N.~Nishizawa, ``Investigating accuracy
  of pitch-accent annotations in neural network-based speech synthesis and
  denoising effects,'' in \emph{Proc. INTERSPEECH}, Hyderabad, Telangana,
  India, Sep 2--6 2018, pp. 37--41.

\bibitem{DumoulinV2016}
V.~Dumoulin and F.~Visin, ``A guide to convolution arithmetic for deep
  learning,'' arXiv:1603.07285 [stat.ML], Mar 23 2016.

\bibitem{IoffeS2015}
S.~Ioffe and C.~Szegedy, ``Batch normalization: Accelerating deep network
  training by reducing internal covariate shift,'' in \emph{Proc. International
  Conference on Learning Representations (ICLR)}, San Diego, CA, USA, May 7--9
  2015.

\bibitem{SchusterM1997}
M.~Schuster and K.~K. Paliwal, ``Bidirectional recurrent neural networks,''
  \emph{IEEE Transactions on Signal Processing}, vol.~45, no.~11, pp.
  2673--2681, Nov 1997.

\bibitem{HochreiterS1997}
S.~Hochreiter and J.~Schmidhuber, ``Long short-term memory,'' \emph{Neural
  Computation}, vol.~9, no.~8, pp. 1735--1780, Nov 15 1997.

\bibitem{ZhangJX2018}
J.-X. Zhang, Z.-H. Ling, and L.-R. Dai, ``Forward attention in
  sequence-to-sequence acoustic modeling for speech synthesis,'' in \emph{Proc.
  International Conference on Acoustics, Speech, and Signal Processing
  (ICASSP)}, Calgary, AB, Canada, Apr 15--20 2018, pp. 4789--4793.

\bibitem{ChorowskiJ2015}
J.~Chorowski, D.~Bahdanau, D.~Serdyuk, K.~Cho, and Y.~Bengio, ``Attention-based
  models for speech recognition,'' in \emph{Proc. Conference on Neural
  Information Processing Systems (NIPS)}, Montr\'eal, QC, Canada, Dec 7--12
  2015, pp. 577--585.

\bibitem{SrivastavaN2014}
N.~Srivastava, G.~Hinton, A.~Krizhevsky, I.~Sutskever, and R.~Salakhutdinov,
  ``Dropout: A simple way to prevent neural networks from overfitting,''
  \emph{Journal of Machine Learning Research}, vol.~15, pp. 1929--1958, Jun 14
  2014.

\bibitem{KruegerD2017}
D.~Krueger, T.~Maharaj, J.~Kram\'ar, M.~Pezeshki, N.~Ballas, N.~R. Ke,
  A.~Goyal, Y.~Bengio, A.~Courville, and C.~Pal, ``Zoneout: Regularizing {RNNs}
  by randomly preserving hidden activations,'' in \emph{Proc. International
  Conference on Learning Representations (ICLR)}, Palais des Congr\`es Neptune,
  Toulon, France, Apr 24--27 2017.

\bibitem{VaswaniA2017}
A.~Vaswani, N.~Shazeer, N.~Parmar, J.~Uszkoreit, L.~Jones, A.~N. Gomez,
  L.~Kaiser, and I.~Polosukhin, ``Attention is all you need,'' arXiv:1706.03762
  [cs.CL], Jun 12 2017.

\bibitem{BahdanauD2015}
D.~Bahdanau, K.~Cho, and Y.~Bengio, ``Neural machine translation by jointly
  learning to align and translate,'' in \emph{Proc. International Conference on
  Learning Representations (ICLR)}, San Diego, CA, USA, May 7--9 2015.

\bibitem{ChoK2014}
K.~Cho, B.~van Merrienboer, C.~Gulcehre, D.~Bahdanau, F.~Bougares, H.~Schwenk,
  and Y.~Bengio, ``Learning phrase representations using {RNN} encoder-decoder
  for statistical machine translation,'' in \emph{Proc. Conference on Empirical
  Methods in Natural Language Processing (EMNLP)}, Doha, Qatar, Oct 25--29
  2014, pp. 1724--1734.

\bibitem{OordA2016}
A.~van~den Oord, S.~Dieleman, H.~Zen, K.~Simonyan, O.~Vinyals, A.~Graves,
  N.~Kalchbrenner, A.~Senior, and K.~Kavukcuoglu, ``{WaveNet}: A generative
  model for raw audio,'' arXiv:1609.03499 [cs.SD], Sep 12 2016.

\bibitem{WangX2018}
X.~Wang, J.~Lorenzo-Trueba, S.~Takaki, L.~Juvela, and J.~Yamagishi, ``A
  comparison of recent waveform generation and acoustic modeling methods for
  neural-network-based speech synthesis,'' in \emph{Proc. International
  Conference on Acoustics, Speech, and Signal Processing (ICASSP)}, Calgary,
  AB, Canada, Apr 15--20 2018, pp. 4804--4808.

\bibitem{TamamoriA2017}
A.~Tamamori, T.~Hayashi, K.~Kobayashi, K.~Takeda, and T.~Toda,
  ``Speaker-dependent {WaveNet} vocoder,'' in \emph{Proc. INTERSPEECH},
  Stockholm, Stockholm, Sweden, Aug 20--24 2017, pp. 1118--1122.

\bibitem{ITU2005}
{International Telecommunication Union}, Recommendation {G.191}: Software Tools
  and Audio Coding Standardization, Nov 11 2005.

\bibitem{Brunner2000}
E.~Brunner and U.~Munzel, ``The nonparametric {Behrens-Fisher} problem:
  Asymptotic theory and a small-sample approximation,'' \emph{Biometrical
  Journal}, vol.~42, no.~1, pp. 17--25, Jan 2000.

\bibitem{MoriseM2016}
M.~Morise, F.~Yokomori, and K.~Ozawa, ``{WORLD}: A vocoder-based high-quality
  speech synthesis system for real-time applications,'' \emph{IEICE
  Transactions on Information and Systems}, vol. E99-D, no.~7, pp. 1877--1884,
  Jul 1 2016.

\end{thebibliography}

\EOD

\end{document}